\documentclass[prl,aps,twocolumn,superscriptaddress,fleqn]{revtex4-2}

\usepackage{
	graphicx,
	subfigure,
	placeins
}
\graphicspath{{Figures/}}

\usepackage{graphicx,subfigure,placeins,wrapfig}

\usepackage[utf8]{inputenc}
\usepackage[T1]{fontenc}
\usepackage{xcolor,booktabs,color,lipsum,makecell,microtype,fix-cm,multirow}
\usepackage[normalem]{ulem}
\usepackage{bm, amsmath,float,amssymb,upgreek,physics,braket,mhchem}
\usepackage[]{siunitx}

\usepackage{nicefrac} 			

\usepackage{tikz}

\DeclareRobustCommand\circled[1]{\tikz[baseline=(char.base)]{\node[shape=circle,draw,inner sep=.6pt] (char) {\scriptsize #1};}}

\usepackage{}

\usepackage{chngcntr,printlen}
\usepackage[unicode=true,colorlinks=true,citecolor=blue,urlcolor=blue]{hyperref}

\begin{document}

\title{Circular  THz ratchets in a 2D-modulated  Dirac system}
	
		\author{M. Hild}
	\affiliation{Terahertz Center, University of Regensburg, 93040 Regensburg, Germany}
		
	\author{I. Yahniuk}
	\affiliation{Terahertz Center, University of Regensburg, 93040 Regensburg, Germany}

	\author{L. E. Golub}
\affiliation{Terahertz Center, University of Regensburg, 93040 Regensburg, Germany}

\author{J. Amann}
\affiliation{Terahertz Center, University of Regensburg, 93040 Regensburg, Germany}

\author{J. Eroms}
\affiliation{Terahertz Center, University of Regensburg, 93040 Regensburg, Germany}

\author{D. Weiss}
\affiliation{Terahertz Center, University of Regensburg, 93040 Regensburg, Germany}

\author{K. Watanabe}
\affiliation{Research Center for Electronic and Optical Materials, National Institute for Materials Science, 1-1 Namiki, Tsukuba 305-0044, Japan}

\author{T. Taniguchi}
\affiliation{Research Center for Materials Nanoarchitectonics,  National Institute for Materials Science, 1-1 Namiki, Tsukuba 305-0044, Japan }

\author{S. D. Ganichev}
\affiliation{Terahertz Center, University of Regensburg, 93040 Regensburg, Germany}

\begin{abstract} We report on the observation of the circular ratchet effect excited by terahertz laser radiation in a specially designed two-dimensional metamaterial consisting of a graphene monolayer deposited on a graphite  gate patterned with an array of triangular antidots. We show that a periodically driven Dirac fermion system with spatial asymmetry converts the a.c. power into a d.c. current, whose direction reverses when the radiation helicity is switched. The circular ratchet effect is demonstrated for room temperature and a  radiation frequency of 2.54\,THz. It is shown that the ratchet current magnitude can be controllably tuned by the patterned and uniform back gate voltages. The results are analyzed in the light of the developed microscopic theory considering electronic and plasmonic mechanisms  of the ratchet current formation.
\end{abstract}

\maketitle

\section{Introduction}
\label{introduction}

Single field-effect transistors (FETs) have proven to be promising devices for sensitive and fast room-temperature detection of terahertz (THz) radiation,  see, e.g., \,\cite{Knap2009,Lisauskas2009,Vicarelli2012,Koppens2014,Otsuji2015,Marczewski2018,Bandurin2018,Rogalski2019,Lewis2019}. In the last decade, it has been suggested that the performance of FET structures can be significantly improved by fabricating an asymmetric comb-like dual grating-gate (DGG) FETs based on semiconductor quantum wells or graphene\,\cite{Olbrich2009ratchet,Olbrich2011,Kiselev2011,Nalitov2012, Otsuji2013,Drexler2013,Kurita2014,Budkin2014, Faltermeier2015, Olbrich2016,Popov2016,Fateev2019,Hubmann2020,BoubangaTombet2020,DelgadoNotario2020,Moench2022, Yahniuk2022, Tamura2022, Moench2023a, Moench2023b, Abidi2024,Yahniuk2024arxiv}. Furthermore, it has been shown that electronic and plasmonic ratchet effects in DGG-based metamaterials lead to  a helicity-driven photoresponse that reverses the sign by switching from right-handed to left-handed circular polarization\cite{Olbrich2009ratchet,Ivchenko2011,Nalitov2012,Faltermeier2015,Rozhansky2015,Olbrich2016,Moench2022,Moench2023a}. This is in contrast to the single FET in which a helicity-driven photoresponse can only be obtained due to an interference of plasma oscillations in the channel of the FETs connected to specially designed antennas\,\cite{Drexler2012,Romanov2013,Gorbenko2018}, e.g., a tilted bow-tie antenna\cite{Matyushkin2020,Matyushkin2021}. Therefore, metamaterials can not only improve the performance of FET detectors, but also lead to new functionalities, in particular, the all-electric detection of  radiation helicity. Thus, the search for novel concepts and  designs of metamaterials that provide the helicity-sensitive dc current in response to terahertz radiation is an important and challenging task. Here, we report on the observation and study of the helicity-driven dc current excited by THz radiation in a  2D metamaterial consisting of a graphite  gate patterned with an array of triangular antidots and placed under a graphene monolayer. We show that the photoresponse is due to the ratchet current caused by the combined action of a spatially periodic in-plane electrostatic potential and a periodically modulated radiation electric field caused by near-field  diffraction. We investigate an all-electrically tunable magnitude of the rectified voltage that is different for clockwise and counter-clockwise circularly polarized radiation.  The data are discussed in the light of the developed theory, which is based on the solution of the Boltzmann kinetic equation and  well describes all experimental findings. The results are analyzed in terms of electronic and plasmonic mechanisms of photocurrent generation in periodic structures. We show that the ratchet photocurrent arises due to the noncentrosymmetric unit cell of  the periodic structure.

\section{Sample and methods}
\label{samples_methods}

\begin{figure}[t] 
	\centering
	\includegraphics[width=\linewidth]{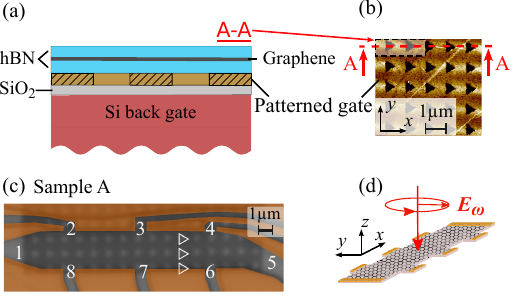}
	\caption{Panel (a): Cross section of the two unit cells, see dashed rectangular in the panel (b). Panel b: (b) AFM image of the patterned gate formed by the  periodic array of triangular antidots arranged in a square lattice. Panel (c): A photograph of sample A (the orange area is the etch mask overlay), in which  triangles are highlighted by a white line and the numbers enumerate the contacts. Panel d: Experimental setup.}
	\label{fig1}
\end{figure}

Figures\,\ref{fig1} (a) and (b) show the design of the  investigated metamaterial. The monolayer  of graphene  encapsulated between hexagonal boron nitride (hBN) is deposited on the patterned  bottom gate made of five layers of graphene.  The pattern consists of  equilateral triangular antidots arranged in a square lattice; see Fig.\,\ref{fig1} (b). The antidots array has a period of \SI{1000}{nm}. In addition, a Si wafer with a   285~nm SiO$_2$ layer on top was   used as a uniform back gate. Two Hall bar samples with ohmic chromium-gold contacts were fabricated. The length and width of the sample A (B) were  \SI{16}{\micro m} (\SI{7.5}{\micro m })  and  \SI{2.5}{\micro m} (\SI{3.5}{\micro m}), respectively. Consequently, the area of the Hall bar structures $A_\mathrm{s}$ was \SI{40}{\micro m^2} for sample A and \SI{26}{\micro m^2} for sample B.
The maximum charge densities for the studied patterned gate voltage $V_\mathrm{p}$, ranging from \SI{-0.75}{V} to  \SI{0.75}{V} were $4 \times 10^{11}$\,cm$^{-2}$ for electrons and  $5.5\times 10^{11}$\,cm$^{-2}$ for holes.   The maximum carrier mobility for electrons and holes are $10^4$\,cm$^2$/Vs and  $0.7 \times 10^4$\,cm$^2$/Vs, respectively. Further details on sample preparation and transport characteristics can be found in Ref.\,\cite{Yahniuk2024arxiv}, where the same samples were investigated. 

\begin{figure}[t] 
	\centering
	\includegraphics[width=\linewidth]{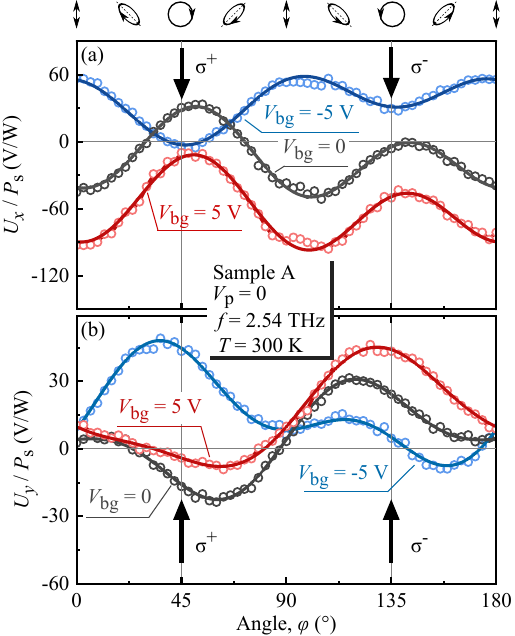}
	\caption{Photovoltage normalized to the radiation power $P_s$ as a function of the angle $\varphi$, which determines the radiation helicity. The data are obtained for sample A. Panels (a) and (b) show the $U_x/P_{\rm s}$ and $U_y/P_{\rm s}$  measured across the long (contacts 1-5) and short (contacts 8-2) sides of the Hall bar for different back gate voltages and for $V_{\rm p}=0$.  Th curves are fits according to Eq.\,\eqref{x-G0} and\,\eqref{y-G0} with the fitting parameters given in Tab.\,\ref{T1}. The ellipses on top illustrate the   polarization states for different angles $\varphi$.	
	}
	\label{fig2}
\end{figure}

To excite ratchet currents, we used polarized THz radiation from a cw molecular laser optically pumped by a CO$_2$ laser with CH$_3$OH as active medium operating at frequency $f = 2.54$\,THz (wavelength $\lambda =118$~$\mu$m) and power at the sample position of about \SI{30}{mW}.  The sample was excited by normally incident elliptically (circularly) polarized radiation, see Fig.\,\ref{fig1}. A Gaussian-like shape of the beam measured by a pyroelectric camera\,\cite{Ganichev1999,Herrmann2016} had a full width at half maximum of $\SI{1.8}{mm}$ and a beam area $A_\mathrm{beam} \approx 0.025$\,cm$^2$  at the sample position. Considering the area of the Hall bar, the sample was exposed to the power $P_\mathrm{s}= P \times A_\mathrm{s}/A_\mathrm{beam}$. The radiation polarization of the initially linearly polarized laser radiation with $\bm E \parallel y$ was obtained using crystal quartz lambda quarter plates. The plate was rotated counterclockwise by an angle $\varphi$ between the $x$-axis and the $c$-axis of the plate. In this geometry, the degree of circular polarization varies according to $P_{\rm circ}=(I^{\sigma^+}-I^{\sigma^-})/(I^{\sigma^+}+I^{\sigma^-})= \sin 2 \varphi$. Here $I^{\sigma^+}$ ($I^{\sigma^-}$) is the intensity of the right- (left-) handed circularly polarized radiation. Consequently, for $\varphi = 45^\circ$ and 135$^\circ$ the radiation  was right-handed  ($\sigma^+$) and left-handed ($\sigma^-$) circularly polarized, respectively.  The polarization ellipses for some angles $\varphi$ are sketched  on top of Fig.\,\ref{fig2}.

The radiation was modulated by an optical chopper at a frequency 130\,Hz and the photovoltage $U$ was measured using the standard lock-in technique. The samples were placed in a vacuum chamber with a $z$-cut crystal quartz optical window covered with a black polyethylene foil to prevent illumination by visible and near infrared light. The voltage $U_x$ was measured at the contacts  5-1 and $U_y$ was measured at the contacts 8-2, with contacts 1 and 2  grounded. The generated ratchet current density was calculated as $j=U/(R_\mathrm{s} w)$, where  $R_\mathrm{s} \ll R_{\rm in}$ is the two-point sample resistance, $R_{\rm in}$ is the input impedance of the amplifier, and $w$ is the Hall bar width. All experiments were conducted at room temperature.

\begin{table}[h]
	
	\centering
	
	\begingroup
	\setlength{\tabcolsep}{5pt} 
	\renewcommand{\arraystretch}{1.4} 
	
	\begin{tabular}{l|crrr}
		\hline
		&Parameter	&	\multicolumn{3}{c}{~$V_\mathrm{bg}$ (V)}	\\
		& (V/W)	& -5 & 0& 5 \\
		\hline 	\hline 
		\multirow{4}{*}{$U_x$}&		$U^\mathrm{circ} $ & 17 &-17 &-25 \\
		&	$U_\mathrm{0} $ & 14 &12 &-31 \\
		&	$U_\mathrm{L1} $ & -41 &53 &59 \\
		&		$U_\mathrm{L2} $ & 5 &-28 &-17 \\
		\hline 	\hline 
		\multirow{4}{*}{$U_y$}&		$\tilde{U}^\mathrm{circ} $ & -19 &20 &-1 \\
		&		$\tilde{U}_\mathrm{0}$  & 25 &4 &-22 \\
		&		$\tilde{U}_\mathrm{L1}$  & 16 &0 &-14 \\
		&		$\tilde{U}_\mathrm{L2}$  & 21 &21 &-3 \\ \hline
	\end{tabular}
	
	\endgroup
	\caption{Parameters used for fitting of experimental curves in Fig.\,\ref{fig2}.}
	
	\label{T1}
\end{table}

\section{Results}
\label{results}

When the structures were illuminated with elliptically polarized radiation, we measured a photosignal $U_x$ from the source-drain contact pair, i.e. via the long side of the Hall bar and the signal $U_y$ perpendicular to it from the contact pair 8-2, Fig.\,\ref{fig1}(c). Figure\,\ref{fig2} shows the photosignal as a function of the angle $\varphi$, which controls the radiation helicity and the degree of linear polarization. The data are presented for sample~A and show that the overall polarization dependence can be well fitted by 
\begin{align} 
	j_x\propto U_x &=  U^{\rm circ} \sin 2\varphi + U_{\rm 0} \nonumber \\  &  -  U_{\rm L1}\, (1+ \cos{4 \varphi})/2 - U_{\rm L2}\,{\sin{4\varphi}\over 2}   \,,\label{x-G0} \\ 
	j_y\propto U_y &= \tilde{U}^{\rm circ} \sin 2\varphi + \tilde{U}_{\rm 0} \nonumber \\  & -\tilde{U}_{\rm L1}\, (1+ \cos{4 \varphi})/2  - \tilde{U}_{\rm L2}\,{\sin{4\varphi}\over 2}   \,. \label{y-G0} 
\end{align}
 The $\varphi$-dependent terms in Eqs.\,\eqref{x-G0},\,\eqref{y-G0} are the Stokes parameters describing the degrees of linear and circular polarization of the elliptically polarized radiation:
\begin{equation}
\label{Stokes}
P_{L1}=-{1+\cos{4 \varphi}\over 2}, \quad P_{L2}=-{\sin{4\varphi}\over 2}, \quad P_{\rm circ}= \sin{2\varphi}.
\end{equation}
As a central result, the figure shows that in all experimental traces obtained for different back gate voltages, the signals excited by left and right circularly polarized radiation are different in magnitude and, in some traces, of opposite sign. This is due to the significant contribution of the circular photoresponse given by the first term on the right-hand side of both equations, which describes the photocurrent proportional to the degree of the circular polarization $P_{\rm circ}$.  Although we obtained similar results for samples~A and B, in the following, we focus on the data obtained in sample~A and present the data for sample~B in  Appendix\,\ref{appendixA}. 

	\begin{figure}[t] 
	\centering
	\includegraphics[width=\linewidth]{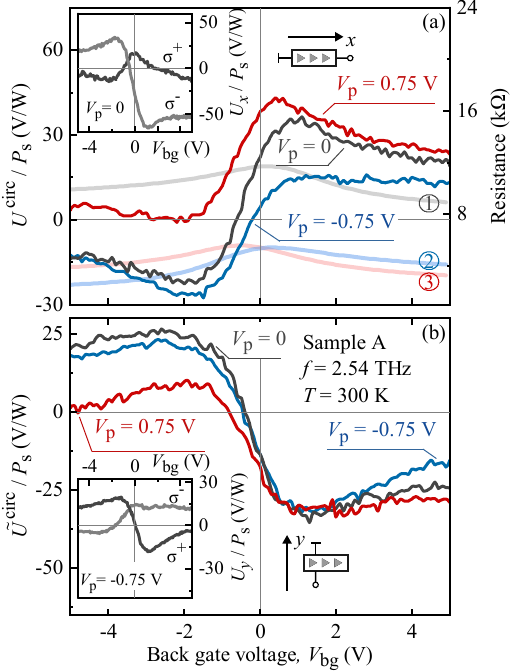}
	\caption{Normalized photovoltages $U^{\rm circ}/P_s$, panel (a) and $\tilde{U}^{\rm circ}/P_s$, panel (b) as a function of the back gate voltage $V_{\rm bg}$. The data are obtained for three values of the  patterned gate voltages $V_{\rm p} = 0, \pm 0.75$\,V. Curves \circled{1}, \circled{2}, and \circled{3} show   back gate voltage dependencies of the two-point  sample resistance (right $y$-axis) obtained for the patterned gate voltages $V_{\rm p}=$ 0, -0.75, and 0.75\,V, respectively. Insets show the back gate dependencies of the signal excited by right  ($\sigma^+$) and left ($\sigma^-$) circularly polarized radiation. Insets show the Hall bar with triangular antidots and contacts used for  signal detection. 
	}
	\label{fig3}
\end{figure}

Figure\,\ref{fig2} and Eqs.\,\eqref{x-G0},\,\eqref{y-G0} show that the helicity-driven photocurrent is superimposed by the polarization-independent ratchet current given by the terms $U_{\rm 0}$ and $\tilde{U}_{\rm 0}$ and the linear ratchet current given by the terms proportional to the coefficients $U_{\rm L1},\tilde{U}_{\rm L1},U_{\rm L2},\tilde{U}_{\rm L2}$. These contributions have been previously detected  and studied using linearly polarized radiation on the same structure, see Ref.\,\cite{Yahniuk2024arxiv}. For pure circularly polarized radiation obtained at $\varphi =45^\circ$ and $135^\circ$,  $P_{L1}$ and $P_{L2}$ vanish. Consequently, the photosignal in response to the circularly polarized radiation is given by
\begin{equation}
U_x^\text{\rm circ} =\pm U^{\rm circ} + U_{\rm 0} \,, \quad U_y^\text{\rm circ} =\pm \tilde{U}^{\rm circ} + \tilde{U}_{\rm 0} \,,
\end{equation}
where the $+$ and $-$ signs correspond to the right- and left- handed circularly polarized radiation respectively. 

\begin{figure}[t] 
	\centering
	\includegraphics[width=\linewidth]{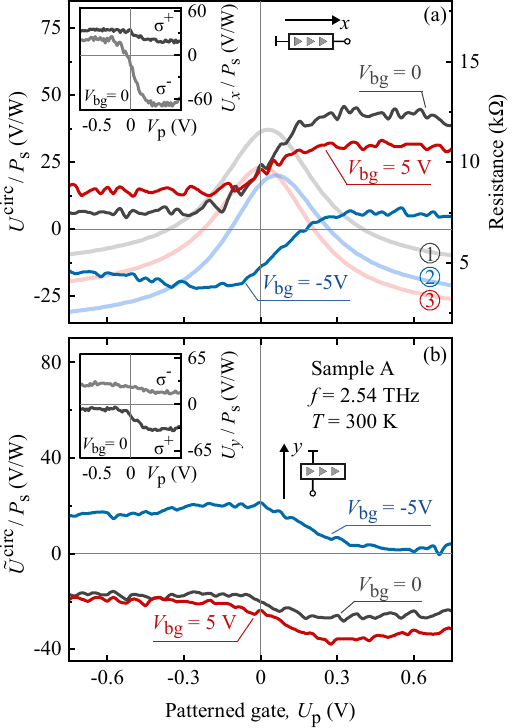}
	\caption{Normalized photovoltages $U^{\rm circ}/P_s$, panel (a), and $\tilde{U}^{\rm circ}/P_s$, panel (b), as a function of the patterned gate voltage $V_{\rm p}$. The data are obtained for three values of the  back gate voltages $V_{\rm bg} = 0, \pm 5$\,V. Curves \circled{1}, \circled{2}, and \circled{3} show the patterned  gate voltage dependencies of the two-point  sample resistance (right $y-$axis) obtained for the back gate voltages $V_{\rm bg}=$ 0, -5, and 5\,V, respectively. Insets show the back gate dependencies of the signal excited by right ($\sigma^+$) and left ($\sigma^+$) circularly polarized radiation. Insets show the Hall bar with triangular antidots and contacts used for  signal detection.}
	\label{fig4}
\end{figure}

 Figure\,\ref{fig2} and the insets in Figs.\,\ref{fig3} and\,\ref{fig4} show that the signals in response to right- and left- handed  circularly polarized radiation are sensitive to the gate voltage: they are consistently different in magnitude, and for certain gate voltages opposite in sign. To analyze the gate voltage dependencies of the helicity-driven photoresponses $U^{\rm circ}$ and $\tilde{U}^{\rm circ}$, we used the fact that these contributions change  sign by reversing the radiation helicity and calculated them according to  $\qty(U_{x,y}^{\sigma^+}-U_{x,y}^{\sigma^-})/2$. 
 
 The gate dependencies of the circular photosignal are shown in Figs.\,\ref{fig3} and\,\ref{fig4}. The curves are obtained by varying one gate voltage $V_{\rm bg}$ or $V_{\rm p}$ and holding   the other constant. Note that due to the patterned graphite gate placed between the graphene layer and the uniform back gate, a voltage applied to the latter gate is in fact not uniform and  introduces an asymmetry of the electrostatic potential acting on the graphene layer. Figures\,\ref{fig3} and\,\ref{fig4} show that varying the asymmetric potential obtained by changing either $V_{\rm bg}$ or $V_{\rm p}$ changes the signal in a similar way: it increases or decreases significantly near the resistance maximum and varies only slightly at high gate voltages. At the same time, depending on the value and sign of the constant gate voltage, it may (i) change  sign near  the resistance maximum, see, e.g., the traces in Fig.\,\ref{fig3}(a) for $V_{\rm p}= -0.75$\,V or Fig.\,\ref{fig3}(b); or (ii) not change sign, see, e.g., Fig.\,\ref{fig4}. A comparison of the results obtained for the source-drain contacts with those obtained for the contact pair across the Hall bar shows that the circular effect for the direction along the  heights of the triangles and their bases   consistently have opposite  signs.

\section{Theory}
\label{theory}

Observation of the photocurrent excited in an unbiased graphene sample using normally incident homogeneous radiation shows that the photoresponse is caused by the asymmetric patterned gate placed under the graphene layer. The asymmetric periodic array reduces the symmetry of the system allowing the generation of such a photocurrent. A previous study of the ratchet effect excited by THz radiation in the same samples demonstrated that  the system under study has $C_1$ point symmetry without any non-trivial symmetry elements\,\cite{Yahniuk2024arxiv}. In the theory, the ratchet current is caused by the simultaneous action of the asymmetric static potential $V(\bm r)$  and the   THz near-field with amplitude ${E}_0(\bm{r})$, formed by radiation diffraction at the edges of the triangular antidots. Both the potential and the near-field are 2D-periodic with the period of the structure. The low symmetry of the studied system is captured by the following vector parameter 
\begin{equation}
	\label{Xi}
	\bm \Xi_{2 \rm D} =\overline{E_0^2(\bm r) \bm \nabla  {V}(\bm r)},
\end{equation}
where the overline indicates the averaging over the 2D period. Although both the near-field intensity and the periodic potential are zero on average, $\bm \Xi_{2 \rm D}$ is finite due to the low symmetry of the structure. 

For structures with $C_1$ point symmetry, both Cartesian components of the vector $\bm \Xi_{\rm 2D}$, $\Xi_x$ and $\Xi_y$, are nonzero. This implies that the ratchet current  allowed under normal light incidence in both $x$- and $y$- directions in the structure plane is:
\begin{align}
	\label{j_C1_x}
	j_x = - \Xi_y  \gamma \:  P_{\rm circ} + \Xi_x \qty(\chi_0  
	+\chi_1 P_{L1}) 
	+\Xi_y
	{\chi}_2 P_{L2},
	\\
	\label{j_C1_y}
	j_y = \Xi_x  \gamma \:  P_{\rm circ} + 
	\Xi_y  \qty(\chi_0 
	- \chi_1P_{L1})+\Xi_x
	{\chi}_2 P_{L2}.
\end{align}
Here $j_{x,y}$ are the components of the electric current density, and the values $\chi_0$, $\chi_{1,2}$ and $\gamma$ describe the polarization independent, linear and circular ratchet currents, respectively. The Stokes parameters of the radiation are defined by the complex polarization vector $\bm e$: $P_{L1}=\abs{e_x}^2-\abs{e_y}^2$, $P_{L2}=e_xe_y^*+e_x^*e_y$, $P_{\rm circ}=-i[\bm e \times \bm e^*]_z$, where we take into account that the radiation propagates along the $-z$ direction. The circular ratchet current reverses its direction when switching from $\sigma^+$ to $\sigma^-$ polarization because it is proportional to the degree of circular polarization  $P_{\rm circ}$. For circularly polarized radiation the circular ratchet current is superimposed on the polarization-independent current, whereas for elliptically polarized light the linear ratchet effect may also contribute, see Eqs.\,\eqref{j_C1_x},\,\eqref{j_C1_y}. As mentioned above, the theory of the polarization independent and linear ratchet currents is presented and  discussed in detail in Ref.\,\cite{Yahniuk2024arxiv}, so in the following, we focus on the theory of the helicity-driven ratchet current.

To develop a microscopic description of the circular ratchet current, we use the drift-diffusion approach based on the solution of the Boltzmann kinetic equation for the distribution function $f(\bm p, \bm r)$
\begin{equation}
	\label{kin_eq}
	\pdv{f}{t} + \bm v_{\bm p} \cdot \bm \nabla f + \bm F(\bm r, t) \cdot \pdv{f}{\bm p} = {\rm St}[f].
\end{equation}
Here $\bm v_{\bm p}  = v_0 \bm p/p$, where  $v_0$ is the Dirac fermion velocity in graphene, ${\rm St}$ is the collision integral, and the force $\bm F$ is a sum of the static contribution due to the potential $V(\bm r)$ and a dynamic force caused by the near field:
\begin{equation}
	\bm F(\bm r, t) = -\bm \nabla V(\bm r) + qE_0(\bm r)[\bm e \exp(-i\omega t) +c.c.],
\end{equation}
where $q$ is the elementary charge positive for holes and negative for electrons. By treating the  periodic potential gradient and the radiation electric field as small perturbations, we iterate the kinetic Eq.\,\eqref{kin_eq} and find the static and homogeneous correction to the distribution function $\delta f(\bm p) \propto \Xi_{x,y}$. This allows us to calculate the ratchet current density as follows:
\begin{equation}
	\bm j = 4 q\sum_{\bm p}\bm v_{\bm p}\delta f(\bm p).
\end{equation}
Remarkably, since we take into account either $\nabla_x V$ or $\nabla_y V$, the final expression for $\bm j$ coincides with the result for the 1D modulation with a potential that depends only $x$ or $y$. Therefore, the 2D character of the modulation is accounted for by the lateral asymmetry parameter $\bm \Xi_{\rm 2D}$, while the factor $\gamma$ in Eqs.\,\eqref{j_C1_x},\,\eqref{j_C1_y} is determined only by the properties of the 2D carriers in graphene above the periodic gate. This problem for $\bm \Xi_{\rm 2D} \parallel x$ has already been  considered for graphene  in Ref.\,\cite{Nalitov2012}. Generalizing to the case of 2D modulation with $\Xi_x\neq 0$ and $\Xi_y\neq 0$ yields, in agreement with the phenomenological Eqs.\,\eqref{j_C1_x}  and\,\eqref{j_C1_y}, the following expression for the circular ratchet current density
\begin{equation}
	\label{j_phenom}
	j_x^{\rm circ} =- \Xi_y  \gamma \:  P_{\rm circ}, \qquad
	j_y^{\rm circ} =\Xi_x  \gamma \:  P_{\rm circ},
\end{equation}
where
\begin{equation}
	\label{gamma}
	\gamma = {q^3 v_0^4 \tau_{\rm tr}^3 \omega^2\mathcal F(\omega\tau_{\rm tr})\over 2s^2 \pi \hbar^2 \varepsilon_{\rm F}[1+(\omega\tau_{\rm tr})^2]
		[\omega^2+(\omega^2-\omega_{\rm pl}^2)^2\tau_{\rm tr}^2]}.
\end{equation}
Here, $\varepsilon_{\rm F}$ is the Fermi energy, $\omega$ is the radiation frequency, and $\tau_{\rm tr}$ is the transport relaxation time. We take here into account a resonant enhancement of the near-field at the plasmon frequency $\omega=\omega_{\rm pl}$\,\cite{Rozhansky2015,Moench2022}. In the studied 2D square lattice with the period $d$, we have $\omega_{\rm pl}=s\sqrt{q_x^2+q_y^2}$, where $q_x=q_y=2\pi/d$, and $s$ is the plasmon velocity.

\begin{figure}[t] 
	\centering
	\includegraphics[width=\linewidth]{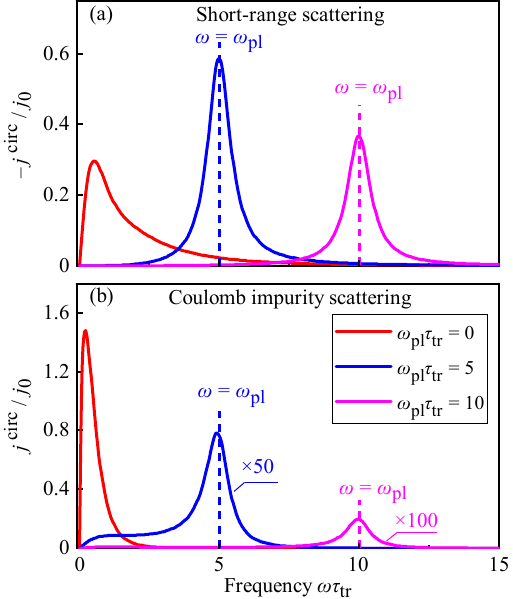}	
	\caption{Frequency dependence of the circular ratchet current for two types of elastic scattering. The curves are calculated for $\omega_{\rm pl}=0$ (red) and for $\omega_{\rm pl}\tau_{\rm tr} \gg 1$, see vertical dashed lines.
	}
	\label{fig_theory}
\end{figure}

The frequency dependence of the ratchet current varies strongly  with the type of the  elastic scattering potential of the carriers in graphene\,\cite{Nalitov2012}. This is captured in Eq.\,\eqref{gamma} by the dimensionless factor $\mathcal F(\Omega)$ where $\Omega=\omega\tau_{\rm tr}$. For the short-range (SR) and long-range Coulomb (Coul) scattering potentials it is given by 
\begin{equation}
	\mathcal F_{\rm SR}(\Omega)=-{\Omega(2\Omega^4+\Omega^2+8)\over(\Omega^2+4)^2},
	\quad
	\mathcal F_{\rm Coul}(\Omega)={1\over \Omega}.
\end{equation}
The divergence at $\Omega \to 0$ that appears in $\mathcal F_{\rm Coul}$ is smeared by  plasmon or energy relaxation processes\,\cite{Nalitov2012}. Therefore, in calculations at $\omega_{\rm pl}=0$ we take $\mathcal F_{\rm Coul} = \Omega/[\Omega^2+(\tau_{\rm tr}/\tau_\varepsilon)^2]$, where $\tau_\varepsilon \gg \tau_{\rm tr}$ is the energy relaxation time. The frequency dependence of the amplitude of the circular ratchet current  generated, e.g., in $y$-direction $j_y^{\rm circ}=\Xi_x \gamma$ normalized to $j_0=\Xi_x q^3 v_0^4 \tau_{\rm tr}^3/(4s^2 \pi \hbar^2 \varepsilon_{\rm F})$ is shown in Fig.\,\ref{fig_theory} for two considered types of elastic scattering potentials. For Coulomb impurity scattering at $\omega_{\rm pl}=0$ we take $(\tau_{\rm tr}/\tau_\varepsilon)^2=0.2$.

\section{Discussion}
\label{discussion}

Now, we discuss the obtained experimental results the light of the developed theory. We begin with the polarization dependence of the observed photoresponse. As discussed in Sec.\,\ref{results}, all curves obtained for different combinations of the gate voltages are well described by Eqs.\,\eqref{x-G0} and\,\eqref{y-G0}, see Fig.\,\ref{fig2}. Comparing these equations with the theoretically obtained Eqs.\,\eqref{j_C1_x},\,\eqref{j_C1_y}   for the ratchet current  {and noting that the Stokes parameters in our experiments vary according to Eq.\,\eqref{Stokes}}, we see that both sets of equations represent identical dependencies on the angle~$\varphi$. The three last terms in each equation with the parameters  $U_0 \propto \Xi_x \chi_0$, $U_{\rm L1} \propto \Xi_x \chi_1$, $U_{\rm L2} \propto \Xi_y \chi_2$, $\tilde{U}_0 \propto \Xi_x \chi_0$, $\tilde{U}_{\rm L1} \propto \Xi_y \chi_1$, and $\tilde{U}_{\rm L2} \propto \Xi_x \chi_2$ represent {the polarization independent} and linear ratchet effects, which are discussed in a separate publication\,\cite{Yahniuk2024arxiv}. Here, we focus on the helicity-sensitive contribution defined by the parameters   $U^{\rm circ}\propto \Xi_y  \gamma$  and $\tilde{U}^{\rm circ}\propto - \Xi_x  \gamma$. Equations\,\eqref{j_phenom}, obtained from both the symmetry arguments and the microscopic theory,  show that the circular ratchet currents excited in $x$- and $y$- direction have opposite directions {provided that $\Xi_x$ and $\Xi_y$ have the same sign}. This is in agreement with the experimental data showing that the corresponding signals have opposite signs, see e.g. Fig.\,\ref{fig2} and insets in Figs.\,\ref{fig3} and\,\ref{fig4}. 

	\begin{figure}[h] 
	\centering
	\includegraphics[width=\linewidth]{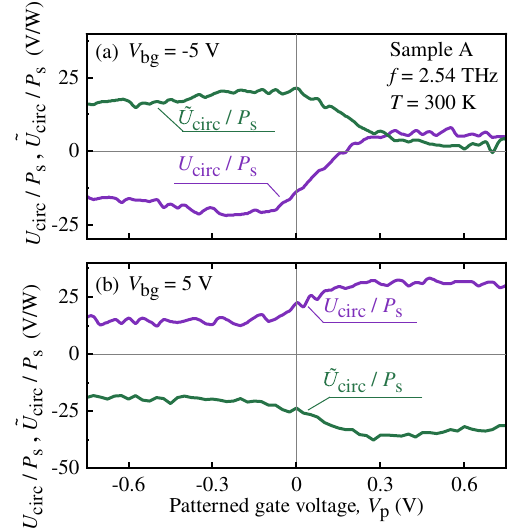}
	\caption{Gate voltage dependencies of the normalized circular photocurrent density $j^{\rm circ}/P_s = U^\mathrm{circ}/(R_\mathrm{s}wP_\mathrm{s})$ measured in the direction along the height of triangles (contacts 1-5). Curves \circled{1}, \circled{2}, and \circled{3} in panel (a) show the back gate voltage dependencies of the two-point  sample resistance (right $y-$axis) obtained for the patterned gate voltages $V_{\rm p}=$ 0, -0.75, and 0.75\,V, respectively. Curves \circled{1}, \circled{2}, and \circled{3} in panel (b) show the patterned gate voltage dependencies of the two-point  sample resistance (right $y-$axis) obtained for the back gate gate voltages $V_{\rm bg}=$ 0, -5, and 5\,V, respectively.
	}	
	\label{fig5}
\end{figure} 

A fingerprint of  the ratchet effects is that it is proportional to the lateral asymmetry parameters $\Xi_x$ and $\Xi_y$, see Eqs.\,\eqref{Xi} and\,\eqref{j_phenom}. These parameters are defined, on the one hand, by the near field, and, on the other hand, by the asymmetry of the electrostatic potential, which can be varied by the gate voltages applied to the back and patterned gates. The simulation  performed in Ref.\,\cite{Yahniuk2024arxiv}  shows that both, patterned and back gate voltages, introduce an asymmetric electrostatic potential acting on the electron gas in graphene. Varying  the gate voltage from negative to positive values changes the magnitude and the sign of the asymmetry. Note that even for zero potential at both gates, an asymmetry is created by the built-in potential caused by the conducting patterned gate deposited under the graphene. The calculations show that the asymmetry along the $x$-direction is indeed present and can be reversed by reversing the polarity of one of the gates. In particular, in the discussed metamaterial the uniform bottom gate potential leads to a change of $\bm \nabla V$ and thus of the lateral asymmetry parameter. This is because the applied back gate voltage is periodically screened by the graphite gate. It can even lead to the formation of lateral $p-n$ junctions.

 Figures\,\ref{fig3} and\,\ref{fig4} show that in the range of $V_{\rm bg}$ from -2 to 2\,V and $V_{\rm p}$ from 0.1 to 0.3\,V the signal measured across the long (short) side of the Hall bar  increases (decreases) linearly with the gate voltages. At higher $|V_{\rm bg}|$ and $|V_{\rm p}|$ the signal saturates or even slightly decreases, see Figs.\,\ref{fig3} and\,\ref{fig4}. The behavior of the ratchet current calculated according to $j=U/(R_\mathrm{s} w)$ is slightly different. This is due to the reduction of the sample resistance with respect to its value at zero gate voltages, see  curves \circled{1}, \circled{2}, and \circled{3}   in Fig.\,\ref{fig3}(a) and Fig.\,\ref{fig4}(a), which affects the gate voltage dependence of the current, especially at high gate voltages.  The gate dependence of the photocurrent is shown in Fig.\,\ref{fig5}, which demonstrates that its behavior only slightly differs from that of the photovoltage: in most cases it saturates at high gate voltages.
 
 Now, we comment the sign of the photocurrent. At first glance, one could expect, since the asymmetry of the electrostatic potential reverses its sign near zero gate voltages, the current should also reverse its direction. However, the real situation  is more complex. Equation\,\eqref{gamma} shows that the current is proportional to the third power of the carriers' charge; thus, varying the gate voltage may also cause the sign of the current to change. Consequently, the reversing of the sign of the electrostatic potential does not necessarily  reverse the ratchet current direction. This is indeed observed in the experiment, which shows that for some values of   one fixed gate voltage, the variation of the other changes the current sign,  while for some  values of the fixed gate the current can keep its direction, see Figs.\,\ref{fig3}(a) and Fig.\,\ref{fig4}. Such a behavior is specific to the used design of our 2D metamaterial, where the patterned gate is placed between the graphene layer and the uniform back gate. Because of the triangular antidots, both gates change the carrier density and  the carrier sign inhomogeneously, either in the area outside the antidots ($n,p$ controlled by $V_{p}$) or above the antidots  ($n^*,p^*$ controlled by $V_{bg}$), the 2D map of the sample resistance showing the gate voltage regions with dominating $n,p,n^*,p^*$ carrier types is presented in Ref.\,\cite{Yahniuk2024arxiv}. By the same arguments, the lateral asymmetry parameters in these two regions, $\bm \Xi_{\rm 2D}$ and $\bm \Xi_{\rm 2D}^*$, can also be different in magnitude and even in sign.  Since the  sign of the ratchet current is defined by the product $q^3 \Xi_{x,y}$ and/or $q^{*3} \Xi_{x,y}^*$,  if both $q$ and $\bm \Xi_{\rm 2D}$ change sign, then the direction of the ratchet current remains unchanged. Otherwise, if one of the signs holds and the other reverses,  then the current direction changes. 
 
Finally, we discuss the frequency dependence of the circular ratchet current, which was obtained theoretically, see Eqs.\,\eqref{j_phenom},\,\eqref{gamma} and Fig.\,\ref{fig_theory}, but has not been studied experimentally so far. The figure shows that the behavior of the circular ratchet current depends strongly on the value of $\omega_{\rm pl}\tau_{\rm tr}$. This is caused by different mechanisms of the ratchet current formation, electronic and plasmonic, realized in the limits $\omega_{\rm pl}\tau_{\rm tr} \ll 1$ and $\omega_{\rm pl}\tau_{\rm tr} \gg 1$, respectively. The frequency dependence for the first case is shown by the  red curves in Fig.\,\ref{fig_theory}. It can be seen that the circular ratchet current behaves non-monotonically with frequency for both types of scattering potentials and has a maximum at $\omega\tau_{\rm tr} \lesssim 1$. This behavior is characteristic for all circular photocurrents caused by the photogalvanic effect, the dynamic Hall effect, and edge photocurrents, for reviews see\,\cite{Glazov2014,Ganichev2017}, and has the same physical background.  At zero frequency, the circular polarization is absent, and subsequently circular photocurrents disappear. At finite frequency, model considerations show that a retardation of  carrier motion with respect to the radiating electric field is crucial for the formation of helicity-driven photocurrents. The retardation is caused by carrier scattering, and the circular photocurrent approaches maximum  
{at $\omega\tau_{\rm tr} \lesssim 1$ for both types of scattering, in particular at $\omega\tau_{\varepsilon} \approx 1$ for Coulomb impurity scattering}.
At higher frequencies, the current  decreases rapidly. Depending on the scattering potential, the  asymptote at $\omega\tau_{\rm tr} \gg 1$ is   $\propto 1/\omega^3$ for short-range scattering but a much faster decrease $\propto 1/\omega^5$ is realized for long-range Coulomb impurities.

Increasing  the plasmon frequency leads  to the plasmonic ratchet effect, which drastically changes the frequency dependence of the circular ratchet current, see blue and magenta curves in Fig.\,\ref{fig_theory} obtained for different values of $\omega_{\rm pl}>1/\tau_{\rm tr}$. For both scattering potentials, in common is that the maximum of the circular ratchet current shifts to frequencies close to $\omega_{\rm pl}$. The width of the plasmon resonance is controlled by the quality factor given by  $\omega_{\rm pl}\tau_{\rm tr}$. For short-range scattering, the amplitude of the current at the resonance is comparable for different values of $\omega_{\rm pl}$. In  sharp contrast, for the Coulomb impurity scattering, an increase of $\omega_{\rm pl}$ results in a drastic decrease of the  current amplitude at  resonance\,\footnote{Note that in the  consideration above we do not take into account effects of inter-particle scattering assuming the corresponding characteristic time is longer than the transport relaxation time due to elastic scattering $\tau_{\rm tr}$. In the opposite case of strong carrier-carrier scattering, the hydrodynamic regime is realized, where  the behavior of the circular ratchet current at high frequencies changes substantially\,\cite{Moench2022}.}. Therefore, it is important to know which type of  scattering potential is responsible for the  ratchet current. The fact that the direction of the current is opposite for  two types of scattering potentials may be helpful in judging this, see Fig.\,\ref{fig_theory}. 

To conclude the discussion on the frequency dependence of the circular ratchet current, we emphasize that the plasmonic ratchet effect for short-range scattering\,\footnote{It is usually realized in small-sized graphene samples}  allows for designing a helicity-sensitive photodetector for the frequency range on demand, because the plasma frequency can be tuned in a wide range by the period of the periodic array.

\section{Summary}
\label{summary}

In summary, experimental results and developed theory show that excitation of  graphene-based 2D metamaterials by circularly polarized THz radiation results in the helicity sensitive ratchet current. In this proof-of-principle work, the circular ratchet current is demonstrated for devices at room temperature and radiation frequency of 2.54~THz. At the same time, the developed theory and previous work on ratchet currents in 1D graphene-based metamaterials\,\cite{Moench2022,Moench2023b} guarantee that such structures should also be efficient over a wide range of temperatures and frequencies. Consequently,  the proposed design of the metamaterial can be considered as a perspective for the development of novel wide-band room-temperature THz detection. As a prospect, the results of the developed theory show that metamaterials with different lattice period and cell size, characterized by different plasmon frequencies, can be used for the development of resonant helicity-sensitive terahertz detectors with the desired central frequency and enhanced responsivity.

\section{Acknowledgments}
\label{acknow}
 
We acknowledge the financial support of the Deutsche Forschungsgemeinschaft (DFG, German Research Foundation) via Project-ID 448955585 (Ga501/18), Project-ID  314695032 – SFB 1277 (Subproject A09),  Project-ID 426094608 (ER 612/2-1), and of the Volkswagen Stiftung Program (97738). K.W. and T.T. acknowledge support from the JSPS KAKENHI (Grant Numbers 20H00354 and 23H02052) and World Premier International Research Center Initiative (WPI), MEXT, Japan.

\appendix
\counterwithin{figure}{section}
\setcounter{figure}{0}

\section{Results obtained on sample B}
\label{appendixA}

\begin{figure}[t] 
	\centering
	\includegraphics[width=\linewidth]{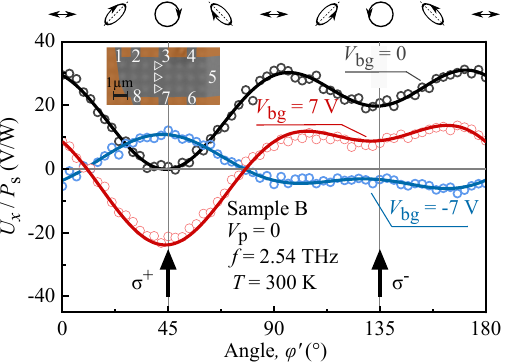}
	\caption{ Normalized photovoltage $U_x/P_s$ measured in sample B (contacts 1-5) as a function of the angle $\varphi$, which determines the radiation helicity. The data are obtained for three values of the back gate voltage $V_{\rm bg}$ holding patterned gate at zero bias. Curves are fits after Eq.\,\eqref{x-G0sampleB} with fitting parameters given in Tab.\,\ref{T2}. The inset shows a photograph of sample B (the orange area is the etching mask overlay). Note that, in contrast to the measurements performed on sample A the Hall bar is oriented vertically, i.e. at zero angle $\varphi$ the radiation electric field is parallel to the $x$-axis.	The ellipses on top illustrate the states of polarization for various angles $\varphi$.
	}
	\label{figA1}
\end{figure}

Here we present the data obtained on the sample B, for its design and structure see the inset in Fig.\,\ref{figA1} and Sec.\,\ref*{samples_methods}. Figure\,\ref{figA1} show the variation of the photosignal magnitude with the angle $\varphi^\prime$. Alike in the sample A we obtain substantial contribution of the helicity sensitive photoresponse.

\begin{figure}[t] 
	\centering
	\includegraphics[width=\linewidth]{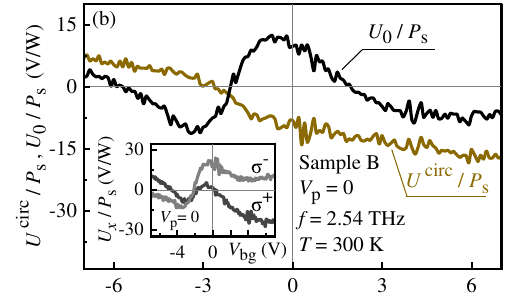}
	\caption{Back gate voltage dependence of the circular $U_{\rm circ}$ and polarization independent  $U_{\rm 0}$ photosignals. The inset shows the back gate dependence of the  photovoltages excited by right- and left-handed circularly polarized radiation.  
	}
	\label{figA2}
\end{figure}

The overall polarization dependence can be well fitted by 
\begin{align} 
	j_x&\propto U_x =  U^{\rm circ} \sin 2\varphi^\prime + U_{\rm 0}  \nonumber \\ 	& +  U_{\rm L1}\, (1+ \cos{4 \varphi^\prime})/2 + U_{\rm L2}\,{\sin{4\varphi^\prime}\over 2}   \,,\label{x-G0sampleB} 
\end{align}
Note that, due to technical reasons in our measurements on sample B, we had to change the definition of the angle $\varphi$: the linear polarization of the laser light was oriented along the $x$-axis, and the angle $\varphi^\prime$ was counted from the $y$-direction. This resulted in the change the Stokes parameters of  the contributions proportional to the coefficients $U_{\rm L1}$ and $U_{\rm L2}$ but didn't affect the contributions $U^{\rm circ}$ and $U_{\rm 0}$. Consequently, for equation for the photoresponse excited by circularly polarized radiation Eq.\,\eqref{x-G0circ} remains unchanged. The inset in Fig.\,\ref{figA2} and the main panel of Fig.\,\ref{figA2} show the back gate dependencies of the signal in response to the right- and left-handed circularly polarized radiation (inset) and of the circular and polarization-independent photoresponses (main panel). It is seen that $U^{\rm circ}$ varies monotonically upon increase of the gate voltage and changes its sign close to the resistance maximum. More complex gate dependence of the polarization independent part is discussed in Ref.\,\cite{Yahniuk2024arxiv}.

\begin{table}[h]
	
	\begingroup
	\setlength{\tabcolsep}{5pt} 
	\renewcommand{\arraystretch}{1.4} 
	\begin{tabular}{l|crrr}
		\hline
		&\multirow{2}{*}{Parameter (V/W)}	&	\multicolumn{3}{c}{~$V_\mathrm{bg}$ (V) }	\\ 
		&	& -7 & 0& 7 \\
		\hline
		\multirow{4}{*}{$U_x$}&		$U^\mathrm{circ} $ & 7 &-10 &-16 \\
		&	$U_\mathrm{0} $ & 4 &10 &-7 \\
		&	$U_\mathrm{L1} $ & -7 &19.5 &16 \\
		&		$U_\mathrm{L2                                                                              } $ & 4 &2 &2 \\
		\hline 
	\end{tabular}
	\endgroup
	\caption{Parameters used for fitting of experimental curves in Fig.\,\ref{figA1}. }	
	\label{T2}
\end{table}

\bibliography{all_lib.bib}
\end{document}